\documentclass[preprint,12pt]{elsarticle}
\usepackage{amsmath,amssymb,amsfonts,mathrsfs, mathtools, bm, bbm, dsfont, mathrsfs,epstopdf}
\usepackage[dvipsnames]{xcolor}
\usepackage{makecell}
\usepackage{enumitem}
\usepackage{enumerate}
\usepackage[font=footnotesize]{subfig}
\usepackage{float}
\usepackage{multirow}
\usepackage{siunitx}
\usepackage[version=4]{mhchem}
\usepackage{booktabs,siunitx}
\usepackage{nomencl}
\usepackage{fullpage}
\usepackage[font=small]{caption}
\usepackage[hyphens]{url}
\usepackage[breaklinks, colorlinks]{hyperref}

\usepackage{tkz-euclide,tikzscale}
\usetikzlibrary{arrows,circuits.ee.IEC}
\tikzset{ac source/.style={
  circuit symbol lines,
  circuit symbol size = width 2 height 2,
  shape = generic circle IEC,
  /pgf/generic circle IEC/before background={
    \pgfpathmoveto{\pgfpoint{-0.8pt}{0pt}}
    \pgfpathsine{\pgfpoint{0.4pt}{0.4pt}}
    \pgfpathcosine{\pgfpoint{0.4pt}{-0.4pt}}
    \pgfpathsine{\pgfpoint{0.4pt}{-0.4pt}}
    \pgfpathcosine{\pgfpoint{0.4pt}{0.4pt}}
    \pgfusepath{stroke}
  },
  transform shape
}}

\newcommand{\solidgreenline}{\raisebox{2pt}{\tikz{\draw[-,OliveGreen,line width = 2pt](0,0) -- (4mm,0);}}}

\newcommand{\solidmidblueline}{\raisebox{2pt}{\tikz{\draw[-,RoyalBlue,line width = 2pt](0,0) -- (4mm,0);}}}
\newcommand{\solidmline}{\raisebox{2pt}{\tikz{\draw[-,magenta,line width = 2pt](0,0) -- (4mm,0);}}}

\newcommand{\boya}[1]{{\textcolor{blue}{Boya says: {#1}}}}

\newcommand{\rev}[1]{{\textcolor{black}{#1}}}

\journal{a journal}

\begin{document}

\begin{frontmatter}

\title{Impact of Aviation Electrification on Airports:\\ Flight Scheduling and Charging}

\author{Boya Hou\corref{cor1}%
\fnref{fn1}}
\ead{boyahou2@illinois.edu}
\author{Subhonmesh Bose \fnref{fn1}}
\author{Lavanya Marla \fnref{fn2}}
\author{Kiruba Haran \fnref{fn1}}

\cortext[cor1]{Corresponding author}

\fntext[fn1]
{Department of Electrical and Computer Engineering, University of Illinois at Urbana-Champaign,Urbana, IL, 61801, USA}
\fntext[fn2]
{Department of Industrial and Enterprise Systems Engineering, University of Illinois at Urbana-Champaign, Urbana, IL, 61801, USA}

\begin{abstract}
Electrification can help to reduce the carbon footprint of aviation. The transition away from the jet fuel-powered conventional airplane towards battery-powered electrified aircraft will impose extra charging requirements on airports. In this paper, we first quantify the increase in energy demands at several airports across the United States (US), when commercial airline carriers partially deploy hybrid electric aircraft (HEA). We then illustrate that smart charging and minor modifications to flight schedules can substantially reduce peak power demands, and in turn the needs for grid infrastructure upgrade. 
Motivated by our data analysis, we then formulate an optimization problem for flight rescheduling that incorporates HEA charging considerations. This problem jointly decides flight schedules and charging profiles to manage airport congestion and peak power demands. We further consider mechanisms via which airlines and airports can negotiate HEA assignments using said optimization problem.
Finally, we illustrate the efficacy of our formulation through a case study on the John F. Kennedy International Airport. 
\end{abstract}


\end{frontmatter}

\section{Introduction}
\label{sec:intro}

Commercial aviation produced 915 million tonnes of CO$_2$ worldwide in 2019, responsible for 2\% of all human-induced CO$_2$ emissions from energy consumption, as per the \citet{ATAG}. The carbon footprint of aviation is projected to increase with predicted annual growth of 4.2\% in demand for air travel over 2018-2038, according to the \citet{ICAO}, \rev{with temporary reductions due to COVID-19 (see \citet{grewe2021evaluating})}. The corresponding increase in greenhouse gas emissions will pose a serious threat to the vision of a carbon-neutral future. Electrification has been identified as a potential path to reduce said emissions, e.g., by the \citet{national2016commercial}.

Electrified aircraft are an emergent technology, largely enabled by the development efforts supported by NASA's Advanced Air Transport Technologies program in the United States and similar programs by the respective agencies in the European Union and Asia.
Various aircraft configurations such as turbo-electric, hybrid-electric and all-electric have been proposed and analyzed. By hybrid electric configuration, we mean those airplanes which are propelled partly by electric motor through a battery system and partly by gas turbines through jet fuel. For example, flight performance of parallel turbofan systems has been analyzed in \citet{Gladin2018}. The benefits of a parallel hybrid propulsion system for boosting power during takeoff and climb has been demonstrated in \citet{Lents2016} and \citet{Bertrand2019}. Small electric airplanes for general aviation are already available. Flight demonstrations are underway for the commuter class with planes carrying $<$20 passengers. The next step is to electrify regional airplanes that accommodate 30 - 100 passengers. Various studies such as those in \citet{schafer2019technological,Wroblewski2019} predict that commercial aviation will adopt electric aircraft over the next few decades (in the 2030-2050 time-frame).

To handle the impending electrification of commercial aviation, airports must invest in appropriate charging infrastructure.  Investment into building such infrastructure must be forward-looking and account for plausible growth trajectories of electrification technology. The first aim of this paper is to provide a framework to gauge the energy and power needs of hybrid electric aircraft (HEA) at major airports across the United States (US), \rev{accounting for the schedules of various airlines at the airport}.  As will become clear, we only consider those HEA configurations that are deemed to become viable over the 2030-2050 time frame, according to academic and industrial research. Scheduling of flights at an airport is intimately related to when and how much the electrified aircraft operating these flights can be charged. As we demonstrate, one substantially impacts the other. The second and final aim of this paper is the formulation and analysis of flight rescheduling protocols at airports that takes charging considerations of HEA into account.
That is, we jointly aim to optimize flight arrivals/departures and decide the charging profiles of HEA, aiming to minimize airport congestion and peak electric power demands from HEA. \rev{To do so, we tally the pros and cons of two possible mechanisms for congestion management. The first mechanism allows a central authority such as the airport to centrally solve the rescheduling problem. The second one designs a negotiation strategy, where airlines propose schedules and the airport runs the rescheduling problem as a feasibility check, and the cycle continues till all parties agree. In this paper, we restrict our congestion management interventions to rescheduling protocols at airports at the strategic or planning phase, which account for airport-wide constraints on movements and charging. Similar questions can be studied for tactical or real-time operations; such investigations are relegated to future work.}

Not all flights can be operated using HEA. Energy densities of today's batteries are in the 200-250 Wh/kg range. They are substantially smaller than that of jet fuel with densities of $\sim$13,000 Wh/kg. As a result, the size and the weight of a battery required on board limits the range of an HEA. The battery size also depends on the extent the aircraft relies on electric propulsion as opposed to jet fuel. With plausible configurations of battery energy densities and degree of hybridization, we compute the energy needs for operating HEA on domestic flight paths within continental US in Section \ref{sec:energy}. By switching flights from current schedules that can be operated by HEA, we estimate the increase in annual energy needs at various US airports. Our estimates indicate that accommodating HEA in commercial aviation will require substantial upgrades to the grid infrastructure that powers these airports. \rev{Thus examining the joint questions of flight scheduling and charging capabilities at airports is a first-order question for the adoption of HEA by airlines, as such infrastructure is essential for adoption.}

While flight distance, number of passengers and airplane type largely dictate the energy needs for HEA, \emph{peak power} requirements from the grid on the other hand, depend on the \emph{rate} of charging. Grid components, \rev{such as transformers,} must be sized properly to support such peak rates. In Section \ref{sec:power}, we show that converting feasible aircraft to HEA without modifying schedules, and charging such HEA at constant power levels over their dwell times at airports, can lead to substantial peak power demands. Thus, optimizing charging schedules can significantly lower these peaks at airports.

Flight arrival and departure schedules define the dwell times of flights at airports. For HEA, these schedules put constraints on when and how much an airplane can be charged. Thus, not surprisingly, alterations in schedules of some flights can further shrink peak power requirements over and above that obtained from optimizing charging schedules alone. Our results in Section \ref{sec:power} indeed align with this expectation. 
Charging considerations alone cannot define flight schedules, however. Airlines tailor their requests for flight arrivals and departures to suit passenger demand patterns. As a result, busy airports often witness congestion during peak hours. Congestion leads to flight delays at these airports. Such delays for multi-hop flight paths tend to cascade across airports. \rev{Flight rescheduling for congestion management has been widely studied, e.g., see \citet{zografos_review_2017} for a survey.} 
\rev{Airports in the European Union and level 3 airports in the US (e.g., John F. Kennedy International Airport (JFK), LaGuardia Airport (LGA), and Ronald Reagan Washington National Airport (DCA)) adopt such mechanisms for congestion management, according to \cite{IATAslot}.}
Along the same lines, we formulate a flight rescheduling problem at an airport that accounts for charging considerations of HEA in Section \ref{sec:SAmodel}.
Specifically, we design an optimization model that seeks to jointly minimize the displacements of flights from their requested schedules by airlines and flatten the power profile required to charge the HEA aircraft operating these schedules.  

\rev{Many American airports today do not impose congestion management protocols. This suggests that one can operate airports without such protocols, albeit with increased and frequent travel delays. The introduction of HEA into airline fleet will likely change that paradigm. Installed capacities of charging infrastructure at airports impose \emph{hard constraints} on charging decisions. In general, one cannot frequently exceed grid equipment limitations, without seriously damaging said equipment. We emphasize that this limitation does not arise solely due to the lack of energy availability. Rather, the installed sizes of transformers and electric power lines for the grid dictate the maximum rate at which energy can be delivered. Thus, feasibility of charging schedules must be checked \emph{across} airlines at an airport. In Section \ref{sec:SAmodel}, we propose two congestion management techniques, building on the rescheduling algorithm with HEA charging considerations. In the first protocol, the airport centrally solves the flight rescheduling and charging problem, and suggests which flight paths should switch HEA with conventional aircraft to ensure acceptable schedule adjustments. In the other, an iterative process ensues, where the airport runs the rescheduling algorithm, but asks each airline to submit a revised schedule, and the process continues till schedule readjustments are acceptable to the airlines and are operationally feasible at the airport. We remark that these are only two among possible mechanisms for congestion management with HEA. A more comprehensive study of congestion management protocols with HEA charging considerations, especially for tactical operations is left for future endeavors.}

We run a representative case study of the rescheduling and charging algorithm with HEA charging for the John F. Kennedy International Airport (JFK) in Section \ref{sec:JFK}. Our experiments reveal the importance of jointly considering flight rescheduling and smart HEA charging with reasonable HEA adoption. In particular, we solve the flight rescheduling problem \emph{without} HEA charging considerations and then construct charging profiles for HEA under a constant power charging scheme. Such a construction results in a high peak power demand of 35.9 MW. Enforcing a limit of 20 MW on power drawn for charging, our optimization problem returns a solution that reduces that peak to 14.6 MW with a different flight schedule. Using our simulation framework, we also study how declared capacities at the airport and charging constraints on airplane batteries impact both flight schedules and charging profiles. \rev{By declared capacity, we mean the quantified measure of the ability of an airport to support a certain number of flight movements within a time duration, that may reflect capacities of runways, terminals/gates, ground-crew, etc.}
The results illustrate that airport congestion and charging considerations are inter-dependent and cannot be tackled separately. \rev{We finally compare the two congestion management protocols. Our results illustrate that more flight paths operate HEA when the airport suggests which HEA route should switch to conventional aircraft, rather than when airlines themselves revise their proposed routes operated by HEA and iterate with airport's feasibility check. In other words, given the coupling between charging and scheduling considerations, centralized joint airport and grid congestion management might play a key role in HEA operation--and over the long run--adoption.}

The key contributions of this paper are as follows. First, we present a framework to study the impact of HEA adoption within commercial aviation. To our knowledge, this is the first paper that jointly tackles transportation and grid considerations with HEA. Past works take solely a power systems perspective or a rescheduling (depeaking) perspective result in solutions that do not capture the grnularity of scheduling considerations and  do not satisfy the  Second, we demonstrate that charging schedules for HEA must be optimized carefully to prevent large peak power demands. Lacking such a design, peak power requirements will put unnecessary burdens on the supporting grid infrastructure.
Our third contribution is the flight rescheduling algorithm with HEA charging. To our knowledge, this is the first work that provides a framework for managing airport congestion and HEA charging together. The case study on JFK airport indicates how one cannot disentangle these two problems once HEA are introduced into airlines' schedules.


\subsection{Literature review}

We draw on two lines of work--one that characterizes the capabilities and impacts of electric airplanes, and the other that studies flight rescheduling and de-peaking algorithms. 
In the first line of research, the  most relevant works are those of \citet{Wroblewski2019, gnadt2019technical} that quantify the capabilities of HEA concepts. We focus on retrofitted regional and single-aisle HEA configurations in Table \ref{tab:concept} that academic and industry research deem viable over the next few decades. Leveraging the technology growth scenarios envisioned in these works, we examine the impacts of HEA adoption on airport operations. We remark that aggregate electricity consumption in the US from electric airplanes has been estimated in \citet{schafer2019technological}. These estimates are based on the uniform adoption scenario of a 180-passenger all-electric airplane, studied in  \citet{gnadt2019technical}. 
In contrast, we provide a detailed systematic framework to study airport operations with much more realistic retrofitted HEA.

\begin{table}[hbtp]
\caption{Summary of regional and single-aisle hybrid electric aircraft concepts and research. BSED stands for battery specific energy density.} 
\label{tab:concept} 
\centering    
            \begin{tabular}{p{4.7cm}c p{6cm}}
            \toprule
            Research group & BSED (Wh/kg) & Reference \\
            \midrule
            Boeing-GE SUGAR Volt   & 750 & \citet{bradley2011subsonic,bradley2015subsonic}\\
            Bauhaus   & 1000-1500 & \citet{Pornet2014a}\\
            UTRC   & Not spec. &\citet{national2016commercial}\\
            Airbus   & 800 & \citet{Delhaye2015}\\
            Cambridge  & 750 & \citet{Friedrich2014}\\
            Georgia Tech   & 750 & \citet{national2016commercial}\\
            \bottomrule
            \end{tabular}
\end{table}


In the second category, there is a growing literature on flight scheduling at capacity-constrained airports. See \citet{NERA,CzernyBook,CorolliEtAl,Benlic2018,zografos_review_2017,pyrgiotis2016impact,jacquillat2018roadmap,ribeiro2018optimization,ribeiro2019large} among others.
These papers optimize flight schedules to limit congestion during peak hours at airports to avoid flight delays, the total cost of which in the US has been estimated to be \$33 billion by the \citet{FAAdelay}. We build on these models to jointly optimize schedules of all flights and the charging profiles of HEA at capacity-constrained airports. \rev{In a sense, we introduce a widely studied question for electrified ground vehicles to the domain of electrified aviation--smart charging for peak demand reduction. Benefits of said reduction are well-studied, e.g., in \citet{veldman2014distribution,spencer2021evaluating,pillai2010impacts,garcia2014plug}, that include deferring infrastructure upgrades of electric power distribution/transmission networks and avoiding peak demand charges. Given the scale of HEA energy and power needs, we believe that smart charging of HEA will become similarly important.}


\newcommand{\Fcal}{\mathcal{F}}

\section{Estimating Energy Requirements of HEA}
\label{sec:energy}

Energy needs of operating a flight path or an aircraft route with HEA will depend on the the type of aircraft being electrified and the distances traveled. For flight paths, we focus on short-haul domestic commercial flights that had a dwell time longer than 15 minutes at the originating airport in 2018. We use flight information from the airline on-time performance data from the \citet{BTSontime}. 
The range of an HEA is limited by the size and weight of the battery on-board. Battery technology for electric airplanes is constantly improving. These batteries are characterized by two parameters---its  battery specific energy density (BSED) and its motor factor (MF). BSED, measured in Wh/kg, dictates the weight of the battery required to deliver a given amount of electrical energy.
And, MF defines the ratio of the peak power that can be delivered by the battery and that required by the aircraft. For a specific BSED-MF combination, we utilize the range capabilities of retrofit hybrid electric regional jets and narrow body aircraft from \citet{Wroblewski2019}, reproduced in  Appendix Table \ref{tab:Range}. A specific flight can utilize HEA only if the flight distance is within this range. 


We now formally describe the electrical energy requirement of operating an aircraft's route or path with HEA. Assume that each HEA arrives at an airport with a depleted battery and needs to be charged up to the level required for its next flight. The required energy is calculated as $E = p  \times  d \times b_0$, where $d$ describes the next flight distance in miles, $p$ is the number of passengers and $b_0$ denotes the battery energy usage per passenger-mile\footnote{The energy use per passenger mile gives a first-order estimate of energy required for aircraft of different sizes (within a reasonable range) with close-to-average occupancy rates.}. This calculation assumes that the electrical power drawn from the battery remains roughly constant during different phases of the flight, e.g. taxi, take-off and landing. For each flight in the database from the \citet{BTSontime}, we use the tail number to identify the aircraft type from airplane manufacturer's websites; which in turn, yields the total number of seats on the plane. Throughout this analysis, we uniformly assume that 85\% of all seats are filled in each flight to estimate $p$. This load factor matches the yearly average estimates of the same in the industry, based on the \citet{BTSstatistics}.  
The values of parameter $b_0$ for HEA are adopted from  \citet{Wroblewski2019}, reproduced in Table \ref{tab:Batt} in the Appendix, assuming a battery-pack voltage of 128V. Sizing of such batteries accounts for battery energy consumed during taxi, takeoff, cruise, approach, and landing. For regional jets, we use $b_0$ for ERJ-175 and for single-aisle aircraft, we use that for Boeing 737-700.

To illustrate the calculations through an example,  consider a single hybrid electric retrofit of Embraer ERJ 170-200 aircraft with tail number N178SY. On 05/29/2018, this aircraft operated a 599-mile flight from SFO (San Francisco International Airport, CA) to SLC (Salt Lake City, UT). N178SY arrived at SFO at 17:02 pm and left for SLC at 17:53 pm. For different MF and BSED configurations, Table \ref{tab:OneFlight} records $E$. For this example, a BSED of 500 Wh/kg and and MF of 25\% does not have the range ability to cover 599 miles. As a result, with this BSED-MF combination, N178SY cannot be operated with HEA.
\begin{table}[!ht]
\centering
\caption {Power requirements of aircraft N178SY for flight from SFO to SLC.} 
\label{tab:OneFlight} 
\begin{tabular}{ccc}
\toprule
\multirow{2}{*}{MF (\%)}  & \multicolumn{2}{c}{BSED (Wh/kg)} \\\cmidrule(lr{1em}){2-3}
 &500 & 700 \\
\midrule
12.5  & 1.05 MWh & 1.03 MWh \\
25  & / & 2.14 MWh \\
\bottomrule
\end{tabular}
\end{table}


\subsection{Annual Energy Requirements at US Airports}
The energy demands for individual flights under BSED-MF combinations prove useful in later sections to both analyze and design charging schedules for HEA. In this study, we consider BSED and MF values that are deemed feasible in the 2030-2050 time-frame, according to \citet{bradley2011subsonic,bradley2015subsonic, Pornet2014a, national2016commercial,Delhaye2015} and \citet{Friedrich2014}.\footnotetext{For BSED below 700 Wh/kg, MF $\geq$50\% is deemed impractical to use for any flight distance.} Here, we utilize our calculations to estimate the increase in annual energy demands at major US airports with plausible growth trajectories of HEA technology, \rev{with flight schedules from the BTS dataset for 2018}. For a given BSED-MF, a flight is feasible to be switched to HEA if its flight distance is within the range of its retrofitted hybrid electric variant. This step defines the maximum size of the domestic fleet that gets electrified. \rev{See  \ref{sec:supp.flight} for the total number of flights that can be operated as HEA under various BSED-MF combinations at the major US airports.}

\begin{figure}[hbtp!]
  \centering
  	\subfloat[ATL]{\includegraphics[width=0.32\textwidth]{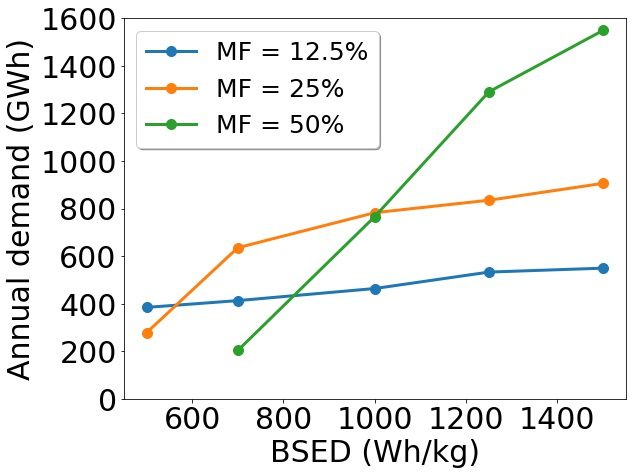}}
	\subfloat[ORD]{\includegraphics[width=0.3\textwidth]{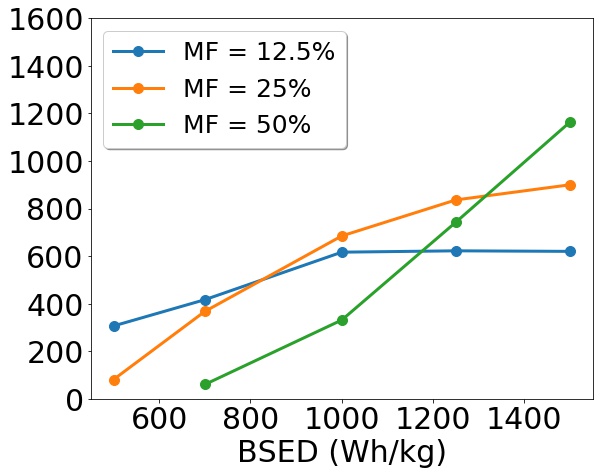}}
	\subfloat[DFW]{\includegraphics[width=0.3\textwidth]{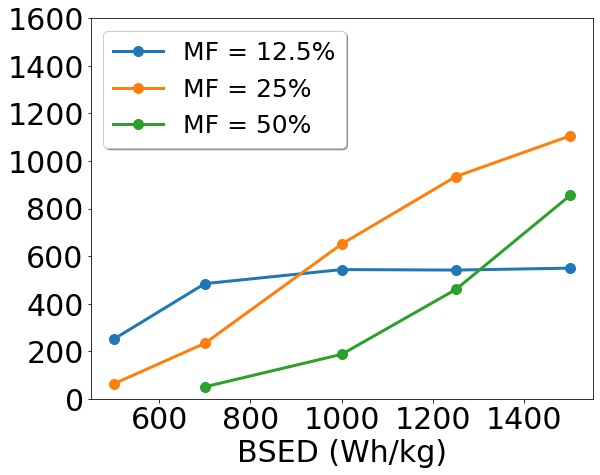}}
	\vspace{0.1in}
	\subfloat[SFO]{\includegraphics[width=0.32\textwidth]{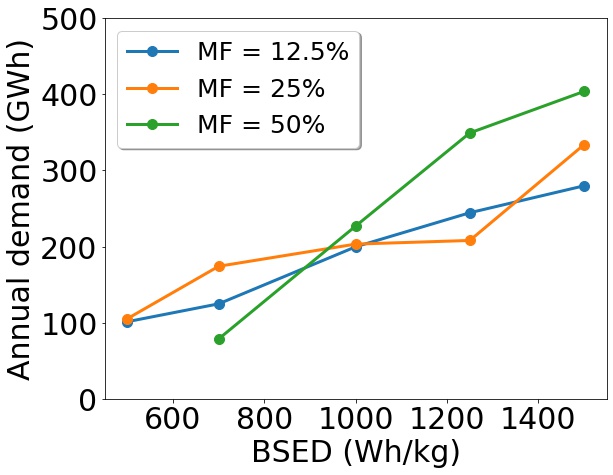}\label{fig:hybrid.SFO}}
	\subfloat[IAD]{\includegraphics[width=0.3\textwidth]{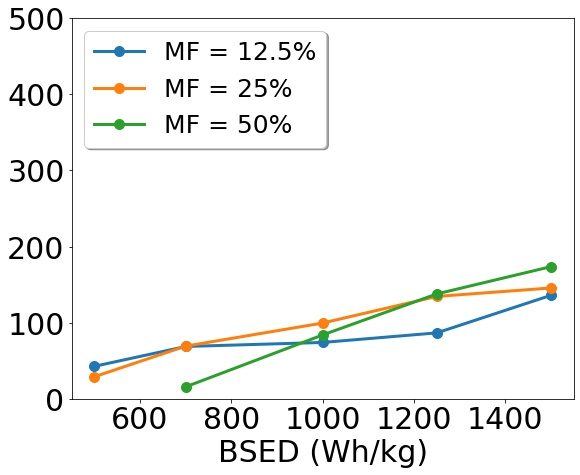}}
	\subfloat[SAN]{\includegraphics[width=0.3\textwidth]{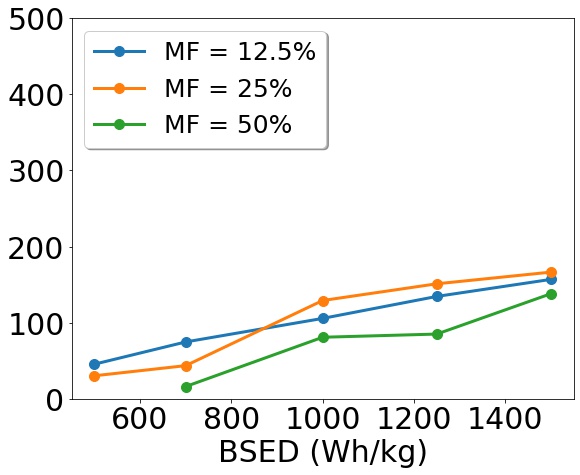}}
	\caption{Extra annual electricity demand that results from charging commercial passenger domestic HEA at airports across US under various BSED-MF combinations.\protect\footnotemark}
  	\label{fig:hybrid}
\end{figure}

Figure \ref{fig:hybrid} plots the projected increase in aggregate annual electricity demands at six large airports in the United States--Hartsfield-Jackson Atlanta International Airport (ATL), Chicago O'Hare International Airport (ORD), Dallas/Fort Worth International Airport (DFW), Dulles International Airport near Washington D.C. (IAD) and San Francisco International Airport (SFO). The plots reveal that even moderate BSED and MF values for HEA will lead to a substantial annual battery energy consumption. To illustrate the magnitude of that increase, notice that aggregate energy demand of SFO in 2018 was 311 GWh, according to the DataSanFrancisco program. Figure \ref{fig:hybrid.SFO} confirms that electrification at SFO with any BSED-MF combination will substantially amplify said demand of 311 GWh. The phenomenon is similar for other airports. For example, ORD had an annual total energy demand of 441 GWh in 2002 according to the O'Hare Modernization Final Environmental Impact Statement. The projected increase in ORD will more than double that requirement even at BSED = 700 Wh/kg and MF = 25\%. For the purposes of this initial analysis, we ignore the possibility that an HEA may need to charge enough to complete a round-trip journey from and to that airport if the destination airport lacks necessary charging infrastructure.  Accounting for such possibilities will only increase our demand estimates.

\begin{figure}[h]
    \centering
    \includegraphics[width=0.4\textwidth]{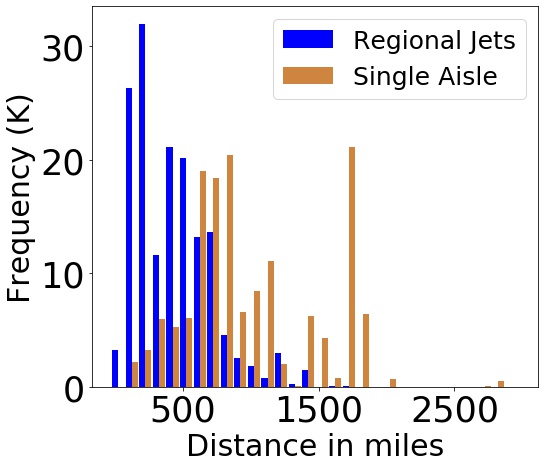}
    \caption{Frequency of flight distances for regional jets and single aisle aircraft flying out of ORD in 2018.}
    \label{fig:distance}
\end{figure}

For a given MF, one might expect total energy consumption from batteries on HEA to decrease with BSED, because one requires lighter batteries to deliver the same amount of power. However, that is not always the trend in Figure \ref{fig:hybrid}. To explain this apparent paradox, we plot the histograms of flight distances in Figure \ref{fig:distance} served by regional jets and single aisle aircraft at ORD in 2018. Notice that the distance distribution of single-aisle aircraft is more right-skewed than that of regional jets. 
Higher BSED values allow larger travel distances. As a result, more single-aisle aircraft, traveling longer distances with higher energy needs, are converted to hybrid. Consequently, energy needs of HEA increase. 


\section{Peak Power Requirements and Peak Shaving Mechanisms}
\label{sec:power}

Grid infrastructure to deliver power at airports must be designed to cover daily peak power demands from HEA and the rest of the airport. In this section, we study the peak power requirements for HEA charging at various airports.

\begin{figure}[h!]
  \centering
	{\includegraphics[width=0.55\textwidth]{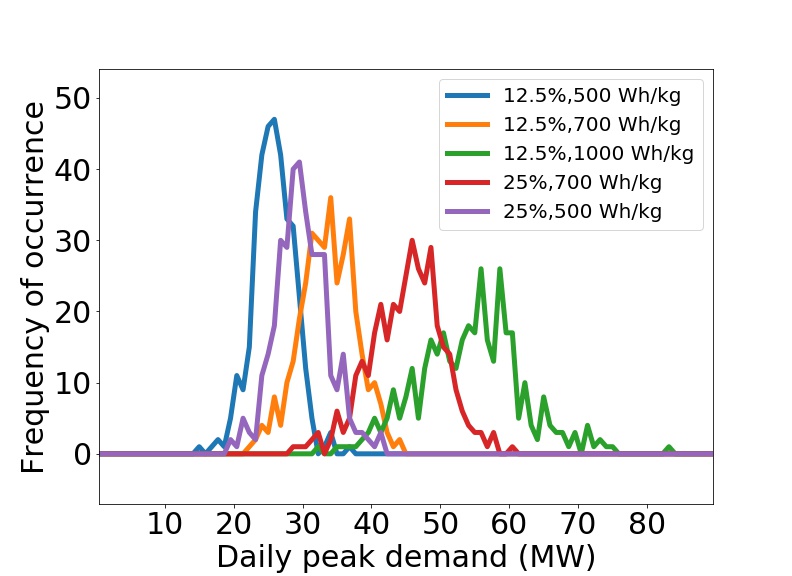}}
	\caption{Histogram of daily peak demands with different BSED (in Wh/kg) and MFs (in \%) at SFO with flight schedule data from 2018.}
  	\label{fig:Hist_Spikes}
\end{figure}

Consider a naive charging scheme, where the energy requirement of HEA is delivered uniformly at constant power over its dwell-time at the airport. This power, summed across all airplanes at each time yields the power requirement of HEA at the airport. In Figure \ref{fig:Hist_Spikes}, we plot the histograms of daily peak powers from HEA charging at SFO using the BTS dataset for 2018 under different BSED settings with MF = 12.5\%. With BSED of 500, 700 and 1000 Wh/kg, we obtain median peak power demands of 25.8, 33.7 and 54.7 MW, respectively. The daily maxima are even higher, e.g., with BSED of 1000 Wh/kg, the highest daily peak is $\sim$82 MW. These demands are substantial, given that the average power demand of SFO in 2018 was 35 MW, according to the DataSanFrancisco program. Supporting grid infrastructure at the airports, including transformers and distribution lines, must be sized to handle the power requirements of HEA. Peak powers from naive charging will pose steep requirements on the grid infrastructure. Even if transformers are properly sized, large peak demands typically increase power procurement costs manifold. This nonlinearity in the growth of procurement costs with peak demand arises due to the fact that generators committed to supply these infrequent peaks have much higher production costs than those used to supply base load. Smart coordinated charging among HEA can shave daily peak power demands.

\subsection{Shaving Peak Power Demands}

We now illustrate the potential of smart charging at airports to reduce daily peak power demands at airports. Assume for this subsection that flight schedules remain the same, that is, the arrival and the departure times for each flight are the same as those in the \citet{BTSontime} database. 
Divide the day into $T=1440$ one-minute intervals. We consider the charging of HEA fleet $\Fcal$, indexed by $n$. For aircraft $n$, let $t_n^A$ and $t_n^D$, respectively, denote its arrival and departure times at the airport gate. Define $E_n$ as its total energy needs for the next flight leg. With $\gamma_n^t$ denoting the charging rate (power) drawn by HEA $n$ in period $t$, we formulate the smart charging problem as 
\begin{alignat}{2} 
\begin{aligned}
& {\text{minimize}} \ \ && \sum_{t=0}^{T-1}\left(\sum_{n \in \Fcal} \gamma_{n}^t\right)^{2}, \\
& \text {subject to} \ \ && 
 \sum_{t =0}^{T-1} \gamma_{n}^t \Delta t = E_{n}, \ \gamma_{n}^t=0 \text{ for } t \notin\left[t_n^A, t_n^D\right], \\
 &&& 
0 \leq \gamma_{n} \leq \overline{Q}_{n}, \ n \in \Fcal,
\end{aligned}
\label{eq:smart.charge}
\end{alignat}
over $\gamma_n^t$ for $n \in \Fcal $ and $t=0,\ldots,T-1$. Here, $\Delta t$ equals 1 minute, the length of the interval. The first constraint enforces HEA $n$ to fulfill its charging obligations over its dwell-time. The second constraint imposes restrictions on charging rates allowed by the airplane battery and the power electronics. Charging power of a battery is often measured in terms of its C-rate. 
A power capacity of 1C \rev{for a specific battery} implies that it requires one hour to fully charge it up to its capacity. \rev{A number $x$C indicates a power capacity $x$ times that of 1C.}
We encode a 10C limit in $\overline{Q}_n$ for each flight, given that higher C-rates are deemed unrealistic (per \citet{fastcharge}), treating the next flight’s energy requirement as the battery capacity\footnotetext{The energy requirement of a flight is upper bounded by the battery capacity. Encoding a C-rate constraint in $\overline{Q}_{n}$ using that capacity ensures that our charging schedule always respects the physical charging rate constraints for batteries.}.
The objective function seeks to flatten aggregate charging profile within the constraints.

\begin{table}[thbp!]
\caption{Highest daily peak power demand in several airports with naive and smart charging.} 
\label{tab:airports} 
\centering    
            \begin{tabular}{cccc}
            \toprule
            \makecell{Airport} &\makecell{Number of flights\\being replaced} &\makecell{ Highest peak (MW)\\ under naive charging}& \makecell{Shaved peak (MW) \\ under smart charging}\\
            \midrule
            ATL&212,205&194.5&98.5\\
            ORD&158,991&138.7&63.5\\
	        DFW&103,177&94.3&38.9\\
            SFO&66,737&60.3&25.2\\
            IAD&35,085&55.1&27.1\\
            SAN&20,876&30.3&14.6\\
            \bottomrule
            \end{tabular}
\end{table}

Table \ref{tab:airports} records the results from six airports for the days with the highest daily peaks from HEA charging under the naive uniform charging scheme using MF=25\%, BSED=700 Wh/kg. The optimization is solved in Python with a Gurobi solver. The results illustrate that daily peaks can substantially reduce through smart charging.



\begin{figure}[hbtp!]
\centering
  	\includegraphics[width=0.5\textwidth]{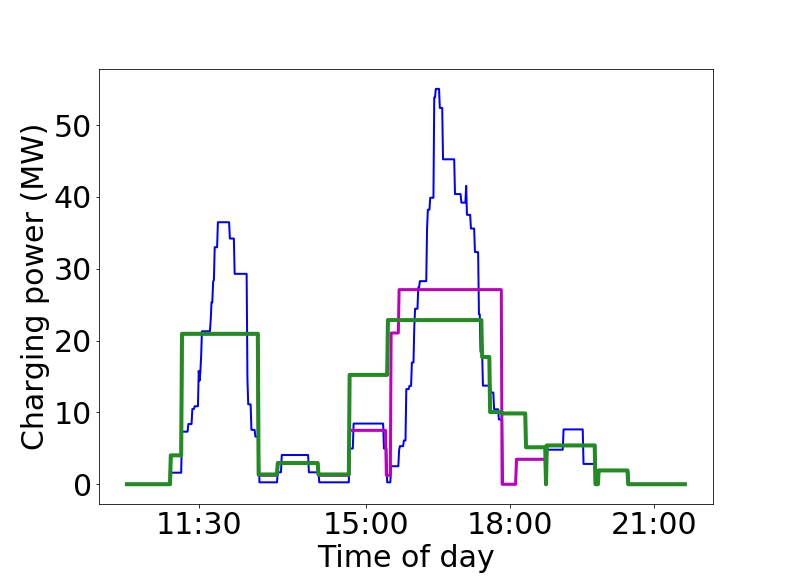}
\caption{Charging profiles with naive uniform charging rate  (\protect\solidmidblueline), smart charging (\protect\solidmline), and smart charging with marginally altered flight schedules (\protect\solidgreenline) at IAD on Dec 6, 2018.}
\label{fig:IAD_30min}
\end{figure}

Marginal alterations in flight schedules can further shave peak power demands. Even a latitude of 30-minutes in the flight departure times of a few flights can reduce daily peaks. For example, the highest daily peak of 55.1 MW for IAD under uniform charging reduces to 27.1 MW under smart charging; even manual marginal flight schedule alteration reduces it further to 22.9 MW. \rev{That is, manually altering the schedules of a few flights within 30 minutes yielded a reduction in peak power from 27.1 MW to 22.9 MW.} Figure \ref{fig:IAD_30min} plots the charging profiles under the three  schemes.

\newcommand{\Scal}{\mathcal{S}}
\newcommand{\Ccal}{\mathcal{C}}
\newcommand{\Hsf}{{\sf H}}

\section{Congestion Management with HEA}
\label{sec:SAmodel}


Our preliminary experiments in the previous section reveal that charging of HEA and scheduling of airplanes must account for the burdens of grid infrastructure upgrades required to support HEA. In this section, we \rev{study congestion management protocols with HEA. We begin in Section \ref{sec:reschedule}  by presenting a flight rescheduling problem that jointly minimizes alterations to submitted schedules by the airlines and charging profiles of HEA to abide by airport operational constraints and charging considerations.
Then in Section \ref{sec:bargain}, we discuss mechanisms for airport congestion control, where airlines propose schedules that include HEA aircraft routes, and airports utilize the flight rescheduling and charging algorithm to check for feasibility and suggest altered schedules. If the output schedule differs too much from an airline's proposed schedule, the airline must update its proposal, based on business needs and operational considerations across multiple airports. One can consider a variety of update mechanisms; we focus on two that allow airlines to alter their aircraft routes operated by HEA to conventional aircraft.}
Altering the usual minimization of \emph{displacements} of output schedules from requested schedules of flights to respect declared capacities at an airport, we tackle charging considerations of HEA within airline fleets. Specifically, we build on  \citet{pyrgiotis2016impact} and \cite{ribeiro2019large} among others, and extend with HEA charging constraints.

\subsection{Flight Rescheduling with Smart HEA Charging Algorithm}
\label{sec:reschedule}
Consider the rescheduling and charging problem over $T$ intervals, denoted $0, \ldots, T-1$. Let $\Scal$ describe the flight scheduling requests (both arrivals and departures) within this horizon; each request is for departure or arrival of a flight in one among $T$ time intervals. Encode the request in
\begin{align}
    A_{i}^t:=
    \begin{cases}1, & \text {if request } i \text{ must be fulfilled no earlier than period } t,
    \\ 
    0, & \text {otherwise }
    \end{cases}
\end{align}
for $t=0, \ldots, T-1$ and $i \in \Scal$.
The binary sequence $\left(A_{i}^0, \ldots, A_i^{T-1}\right)$ assumes the form $(1, \ldots, 1,0, \ldots, 0)$, where the position of the last one  indicates the time interval to execute request $i$. Akin to $A_{i}^{t}$, define $Y_{i}^{t}$ for $i\in\Scal$ and $t=0, \ldots, T-1$ that encodes the allocation decisions instead of requests. That is,
\begin{align}
    Y_{i}^t:=
    \begin{cases}1, & \text {if allocation } i \text{ is fulfilled no earlier than period } t,
    \\ 
    0, & \text {otherwise }
    \end{cases}
\end{align}
for $t=0, \ldots, T-1$ and $i \in \Scal$.
For meaningful allocations, we must have
\begin{align}
    Y_{i}^{t} \geq Y_{i}^{t+1}, \qquad Y_{i}^{1} = 1, \qquad Y_{i}^{t} \in \{0, 1\}
    \label{eq:Y.def}
\end{align}
for all $i \in \Scal$ and $t=0,\ldots,T-1$. 
These constraints imply that $\left(Y_{i}^0, \ldots, Y_{i}^{T-1}\right)$ becomes a sequence of the form $(1, \ldots, 1,0, \ldots, 0)$, where the position of the last one describes the time interval allocated to request $i$. Declared capacity of an airport is described by the number $\overline{R}$ of arrivals and departures that an airport can handle within a horizon of $L$ time intervals. This number encodes operational constraints that arise due to limited number of runways, gate management schemes and available staff, among others.
Thus, we impose the constraint
\begin{align}
\sum_{i \in \Scal} \sum_{\tau=t}^{\min\{ t+L, T-1\}}\left(Y_{i}^{\tau} - Y_{i}^{\tau+1}\right) \leq \overline {R}
\label{eq:runway}
\end{align}
for each $t = 0, \ldots, T-1$.
\rev{In general, many airports consider different capacities over different time horizons; we use a representative $\overline{R}$ for simplicity.}

Let $\Ccal$ describe the set of pairs $(j,j')$ of requests from $\Scal$, where $j$ is an arrival request and $j'$ is the corresponding departure request. Then, we impose a lower bound $\underline{W}_{j,j'}$ on connecting times for flights at the airport as 
\begin{align}
\sum_{t=0}^{T-1}  \left(Y_{j'}^{t}-Y_{j}^{t}\right) \geq \underline{W}_{j,j'}
\label{eq:connect}
\end{align}
for all $(j,j') \in \Ccal$, \rev{given that the intervals $t$ over which $Y_{j'}^t = 1$ and $Y_{j}^t = 0$ are exactly those for which the airplane is at the airport.}

Let $\Ccal^{\Hsf}$ denote the subset of $\Ccal$ with requests of HEA. For $(j,j') \in \Ccal^{\Hsf}$, let $E_{j,j'}$ denote the total energy demand for the aircraft whose arrival/departure requests are indexed by $j,j'$. Let $\gamma_{j,j'}^{t}$ denote the charging rate during time interval $t$ for the  battery of the aircraft that is identified by the requests $j,j'$. The energy needs of that aircraft is enforced via
\begin{align}
\sum_{t=0}^{T-1} \left(Y_{j'}^{t}-Y_{j}^{t}\right) \gamma_{j,j'} ^{t} \Delta t = E_{j,j'}, \qquad \gamma_{j,j'}^{t} \geq 0. 
\label{eq:energy}
\end{align}
Here, $\Delta t$ is the length of each time interval \rev{and hence, $\gamma_{j,j'}^{t} \Delta t$ is the energy delivered to a connecting airplane over time interval $t$.}
Such a constraint is enforced for all $(j, j') \in \Ccal^{\Hsf}$. In addition, we impose two sets of constraints on the power delivered to the HEA. First, the aggregate power for charging all HEA across the airport is constrained by $\overline{P}$, the capacity defined by the grid infrastructure at the airport, as
\begin{align}
\sum_{(j,j') \in \Ccal^{\Hsf}} \left(Y_{j'}^{t}-Y_{j}^{t}\right) \gamma_{j,j'}^{t} \leq \overline{P}
\label{eq:power.airport}
\end{align}
for each $t=0, \ldots, T-1$. Second, we enforce that charging rates for each individual battery does not exceed 10C. Specifically, we impose an upper bound $\overline{Q}_{j,j'}$ on the charging rate of the form
\begin{align}
    \gamma_{j,j'}^{t} \leq \overline{Q}_{j,j'} 
    \label{eq:power.battery}
\end{align}
for each $(j,j') \in \Ccal^{\Hsf}$ and $t=0,\ldots,T-1$. Similar to that in \eqref{eq:smart.charge}, we use the energy requirement of the flight as a proxy for the battery capacity to compute $\overline{Q}_{j,j'}$.


For a request $i \in \Scal$, define its \emph{displacement} as the positive (respectively, negative) difference $X_i^+$ (respectively, $X_i^-$) between the time interval allocated and the time interval requested, i.e.,  
\begin{align}
 X_{i}^{+} := \sum_{t=0}^{T-1}\left(1-A_{i}^{t}\right) Y_{i}^{ t}, \qquad 
 X_{i}^{-} := \sum_{t=0}^{T-1} A_{i}^{ t}\left(1-Y_{i}^{t}\right).
 \label{eq:X.def}
\end{align}

With this notation, we formally present the flight rescheduling and charging problem as the following optimization problem.\footnote{\rev{In the interest of concreteness, we design \eqref{eq:slot.alloc} with a specific objective function. One can formulate the same with additional considerations for both transportation and charging.}} 
\begin{alignat}{2}
\begin{aligned}
& {\text{minimize}} \ \ &&  \max _{i \in \Scal} \max \left\{X_{i}^{+}, X_{i}^{-}\right\} + \sum_{i \in S}\left(X_{i}^{+}+X_{i}^{-}\right)
+ w \sum_{t=0}^{T-1} \left(\sum_{(j,j') \in \Ccal^{\Hsf}} \gamma_{j,j'}^{t}\right)^{2}, 
\\
& \text {subject to} \ \ 
&& \eqref{eq:Y.def} \text{ for } i \in \Scal, t=0, \ldots, T-1, 
\\
&&& \eqref{eq:runway} \text{ for } t=0, \ldots, T-1,
\\
&&& \eqref{eq:connect} \text{ for } (j,j') \in \Ccal, 
\\
&&& \eqref{eq:energy}, \eqref{eq:power.airport}  \text{ for } (j,j') \in \Ccal^{\Hsf},
\\
&&& \eqref{eq:power.battery}  \text{ for } (j,j') \in \Ccal^{\Hsf}, t=0, \ldots, T-1,
\\
&&& \eqref{eq:X.def} \text{ for } i \in \Scal
\end{aligned}
\label{eq:slot.alloc}
\end{alignat}
over the variables $Y$, $\gamma$ and $X$. The objective function is a weighted combination of three terms. The first term is the maximum displacement. The second summand equals the total displacement over all flights. The third summand is a penalty that is designed in a way that minimizing it favors flat aggregate charging profiles of HEA \emph{across} flights, similar in spirit to the smart charging problem in \eqref{eq:smart.charge}. The positive constant $w$ controls the trade-off between minimizing displacements and peak-shaving in charging the HEA. Assigning a low weight $w$ amounts to prioritizing the minimization of displacements of movement requests at the expense of higher peak powers required to charge the HEA. Note that while $\overline{P}$ in \eqref{eq:power.airport} imposes a hard constraint on the total power drawn by HEA at the airport, the third term in the objective function with $w > 0$ seeks to additionally flatten the demand profile within these limits.  $\overline{P}$ encodes capacity constraints of the supporting grid infrastructure at the airport. Operating within these limits, peak shaving is crucial to minimize energy costs of airports. Sharp peaks in power demands are typically met with expensive generators, \rev{the added expense of which are levied on consumers through peak demand charges. These charges are calculated based on the maximum power usage, instead of the net energy consumption, and are common elements of electric utility rate structures across the US. Interaction between electric transit buses and peak demand charges are recorded in \citet{gallo2014peak}.}
Given the magnitude of the peak charging power requirements of airports due to HEA charging, electric peak demand charges can be substantial for airports. Airlines paying for such charges will likely pass these costs on to passengers, increasing travel costs. Positive $w$ can aid in flattening the power profile and reducing the peak power below $\overline{P}$. 




\newcommand{\Acal}{{\mathcal{A}}}
\newcommand{\Xcal}{{\mathcal{X}}}

\subsection{\rev{Alteration of Proposed Schedules Based on Outputs of Rescheduling Algorithm}}
\label{sec:bargain}

\rev{Consider the setting where multiple airlines propose their schedules (including the operation of certain aircraft routes with HEA), but the rescheduling and charging algorithm used by the airport outputs a modified schedule that is quite different from the proposed one. Airlines may find large modifications to their schedules unacceptable, given their business considerations. Therefore, one needs mechanisms for airlines to alter their proposed schedules and iterate the process with the airport. In what follows, we discuss two different alteration procedures and compare their efficacy through our case-study in Section \ref{sec:JFK}. We remark that there is a growing literature on tactical mechanisms for congestion management, e.g., in \cite{mukherjee2009dynamic,guo2022air}. Our focus is on the scheduling stage, and we present two representative among many possible mechanisms. A more comprehensive study of scheduling and tactical congestion management with HEA charging is left for future work.}

\rev{Let $\Acal$ denote the set of airlines. For $a \in \Acal$, let $\Scal_a$ define the set of scheduling requests from airline $a$. Further, let $\Ccal_a$  define the set of pairs $(j, j')$ of requests from $\Scal_a$, where $j$ is an arrival request and $j'$ is the corresponding departure request. Define the subset $\Ccal_a^{\Hsf}$ of $\Ccal_a$ that contains the requests from HEA. Consider the case where all airlines $a \in \Acal$ have submitted their scheduling requests $\Scal_a$ to the airport. Then, the airport runs the rescheduling and charging problem in \eqref{eq:slot.alloc}. If the maximum among the optimal displacements $X_i^+$ and $X_i^-$ from \eqref{eq:slot.alloc} exceed a pre-set limit such as 20 or 30 minutes, some among the airlines must alter their proposed schedules and/or aircraft routes operated via HEA. Our focus being on HEA operation, assume henceforth that an airline can only \emph{switch} an HEA-operated connection to that operated via a conventional aircraft. That is, it only modifies $(j,j')$ from $\Ccal^{\Hsf}_a$ to $\Ccal \setminus \Ccal^{\Hsf}_a$. We consider two different mechanisms for switching, that we describe below.}

\rev{In the first mechanism, the airport chooses which HEA-operated connections must switch to conventional aircraft. Having the information from all airlines, the airport is perhaps the best candidate to make the switching decision. However, such a mechanism takes away an airline's autonomy to decide the aircraft type, and can be practically challenging to implement. To this end, we consider a second mechanism where the airport asks \emph{all} airlines to reduce their positive number of HEA-operated connections at the airport by at least one and resubmit the schedule. We study such a mechanism with the view that an airport cannot discriminate among airlines, and hence, must ask all airlines to make a uniform (or equitable) change to the number of HEA-operated connections. There are a myriad of ways to be equitable amongst airlines for rescheduling, e.g., accounting for the number of passengers affected, size of airlines, etc., for example, see \cite{jacquillat2018interairline} -- however, our focus is not to find the best allocation but to demonstrate the synergy between schedule depeaking and HEA charging. For this reason, we focus on just two such mechanisms, with the viewpoint that one can conceptually consider other potential mechanisms for multi-airline rescheduling with HEA charging.}

\rev{In the first mechanism, the airport can optimize over switching decisions by modifying \eqref{eq:slot.alloc}. Specifically, consider a binary variable $Z_{j,j'} \in \{0, 1\}$ for each $(j,j') \in \Ccal^\Hsf$ and replace the charging rate $\gamma_{j,j'}^t$ during interval $t$ with the product $Z_{j,j'} \gamma_{j,j'}^t$ throughout in \eqref{eq:slot.alloc}. To favor retaining HEA connections, augment the objective with a term $-w' \sum_{(j,j') \in \Ccal^{\Hsf}} Z_{j,j'}$ for a suitable weight $w'>0$, and optimize over $Z$ along with $Y, \gamma$ and $X$. One can then run this optimization problem with sequentially decreasing choices of $w'$ till a schedule that is acceptable to all airlines is found. One can also add a constraint of the form $0 \leq X_i^+, X_i^- \leq \overline{X}$, where $\overline{X}$ captures the maximal displacement that airlines accept.}

\begin{figure}[hbtp!]
\centering
  	\includegraphics[width=0.8\textwidth]{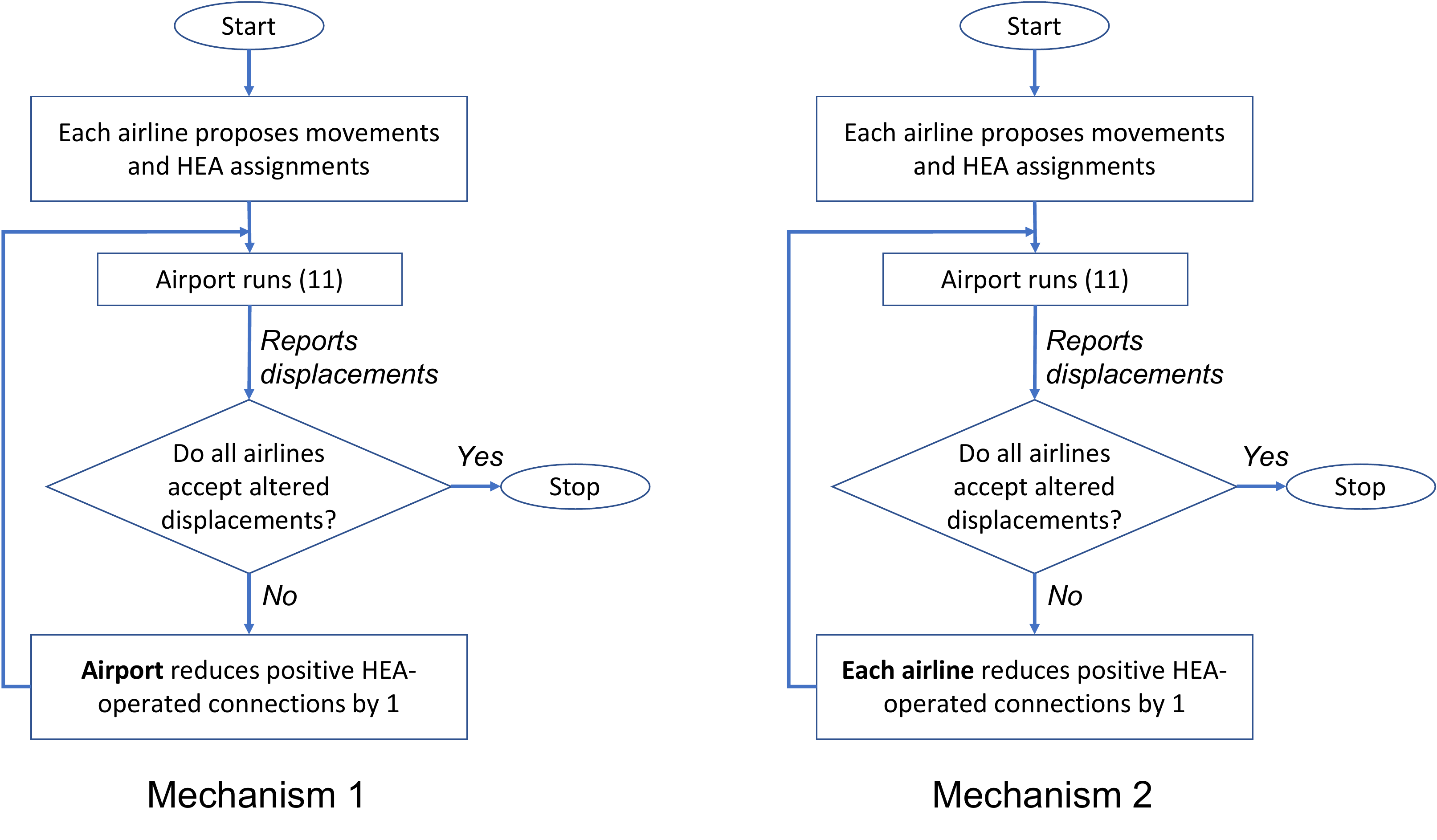}
\caption{Two heuristic mechanisms to alter HEA-operated connections.}
\label{fig:IAD_30min}
\end{figure}

\rev{ Our focus being on the difference between the two mechanisms, we consider an alternate heuristic. Note that any connection whose next flight on the route requires a high \emph{uniform charging rate} over the dwell time at the airport imposes heavy burdens on the charging infrastructure, accommodating which likely causes large displacements of flights. This intuition suggests that the airport might consider ranking the connections with HEA in descending order of the uniform charging rates and then switch them one at a time, till an acceptable schedule is found. If the proposed schedules with all HEA switched to conventional aircraft results in acceptable displacements for all airlines, then this iterative procedure must converge. When the airlines themselves switch the HEA-operated connections, they can adopt the same heuristic, and sort their own HEA-operated connections in decreasing order according to uniform charging rate. Then, each time the airport's rescheduling is deemed unacceptable by any airline, each airline follows this sorted order to remove HEA-operated connections one or more at a time and submits the schedule to the airport. The airport resolves \eqref{eq:slot.alloc} and repeats the procedure with the airlines, if the result is unacceptable. Again, this cycle must end in finitely many iterations, if proposed schedules with all HEA switched to conventional aircraft is acceptable to all airlines.}

\section{Case Study of Flight Rescheduling due to HEA for the JFK Airport}
 \label{sec:JFK}

We now conduct a representative case study for domestic flight operations at the JFK airport based on flight schedules on December 27, 2018. Specifically, we utilize the schedules of domestic flights from the BTS database for the JFK airport as the requested movements over the peak hours of 10:00-16:00, considering each time interval to be 2 minutes in length. Per the BTS dataset, there were 215 movements during this time window. We consider the actual movements as the scheduling requests. Among the requesting aircraft, 32 of them with arriving /departure request pairs can be switched to the hybrid electric option, based on the selection criterion described in Section \ref{sec:energy} with BSED of 700 Wh/kg and MF of 25\%\rev{--conservative parameter choice for attainable BSED and mid-range of expected MF in the 2030-2050 time-frame}. According to the \citet{FAAcap}, JFK supports around 90 arrivals and departures per hour. 
For our first set of experiments, we consider a capacity of $\overline{R} = 45$ over $L = 30$ slots for domestic flights, roughly allocating half the total capacity at JFK to domestic flights, per the share of domestic flights among all flights served at JFK (see the report by \citet{ATR2018}). We encode a minimum connecting time of 30 minutes in $\underline{W}$ for all aircraft. The rescheduling and charging problem in \eqref{eq:slot.alloc} is solved in Python with Gurobi (\citet{optimization2012gurobi}).

\begin{figure}[!hbtp]
  \centering 	
    \includegraphics[width=0.35\textwidth]{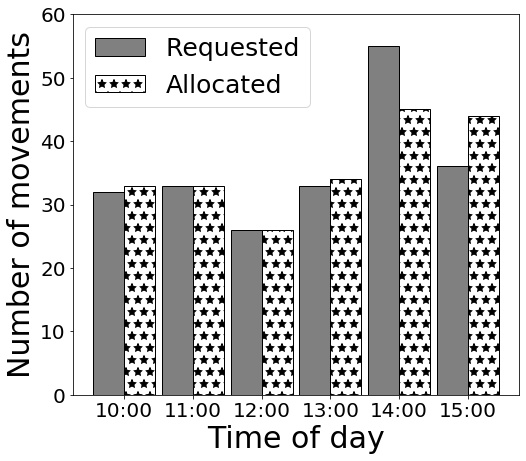}
    \caption{Requested and allocated movements per hour at JFK airport on 12/27/2018;  allocation decisions are based on the benchmark experiment that does not consider HEA charging.}
    \label{fig:comp_slots}
\end{figure}


First, we run the optimization problem in \eqref{eq:slot.alloc} without accounting for charging considerations. That is, we drop the constraints \eqref{eq:energy}, \eqref{eq:power.airport} and \eqref{eq:power.battery} in \eqref{eq:slot.alloc} and set $w = 0$. In effect, we run an optimization problem only over the displacement variables $Y, X$. The outcome of this experiment serves as a benchmark to compare the subsequent results with charging considerations. Figure \ref{fig:comp_slots} illustrates that the resulting allocation decisions exhibit some displacements from requested schedules. Such displacements are inevitable, given that the peak hourly slot request exceeds our assumed airport's considered capacity of $\overline{R} = 45$. With the resulting rescheduling decisions, we construct a charging profile for HEA that will result from a uniform charging rate over their dwell times at the JFK airport. Charging HEA at a constant power level over the resulting allocations then yields  a peak power demand of 35.9 MW.

\begin{figure}[!hbtp]
  \centering   
    \subfloat[Histogram of positive slot displacements  ]{\includegraphics[width=0.46\textwidth]{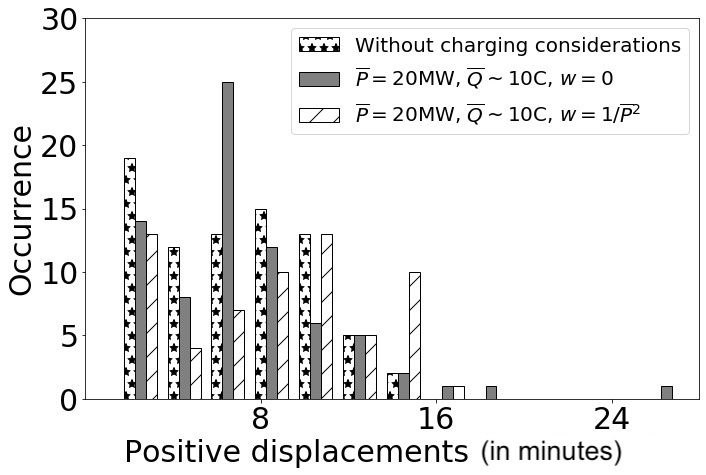}\label{fig:displacements.1}}
    \qquad
  	\subfloat[Charging profiles for HEA]{\includegraphics[width=0.46\textwidth]{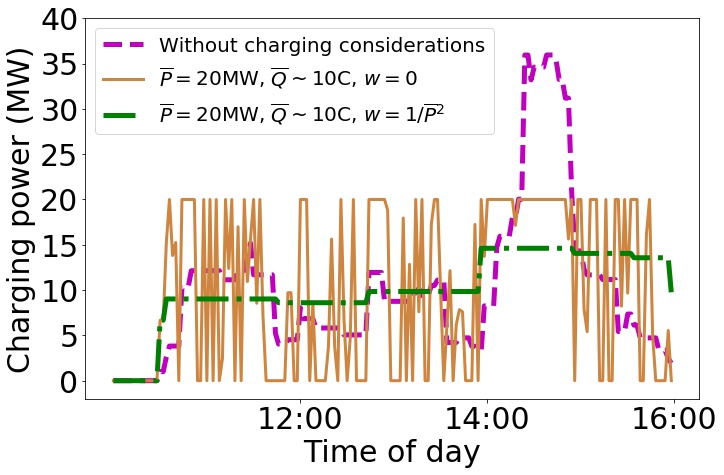}\label{fig:charging.1}}
  	\caption{Positive displacements \protect\footnotemark and charging profiles for three different parameter sets: (1) Rescheduling decisions without charging constraint under uniform charging (2) $\overline{P} = 20$ MW, $w = 0$, $\overline{Q} \sim $ 10C and (3) $\overline{P} = 20$MW, $w = 1/\overline{P}^2$, $\overline{Q} \sim $ 10C. For all experiments, $\overline{R}$ is held constant at 45.}
\end{figure}

\footnotetext{\rev{By positive displacement, we mean non-zero $X_i^{+} + X_i^{-}$, i.e., we omit the cases with no displacement and $X_i^{+} + X_i^{-} = 0$.}} 

\begin{figure}[!hbtp]
  \centering   
    \subfloat[Histogram of positive displacements ]{\includegraphics[width=0.45\textwidth]{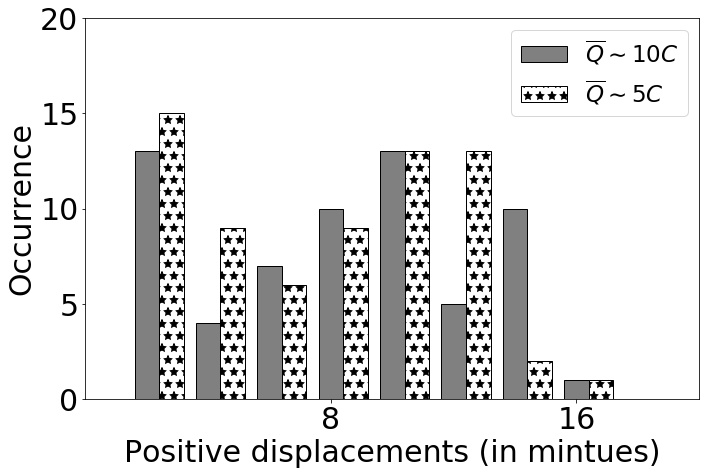}\label{fig:displacements.2}}
    \qquad
  	\subfloat[Charging profiles for HEA]{\includegraphics[width=0.465\textwidth]{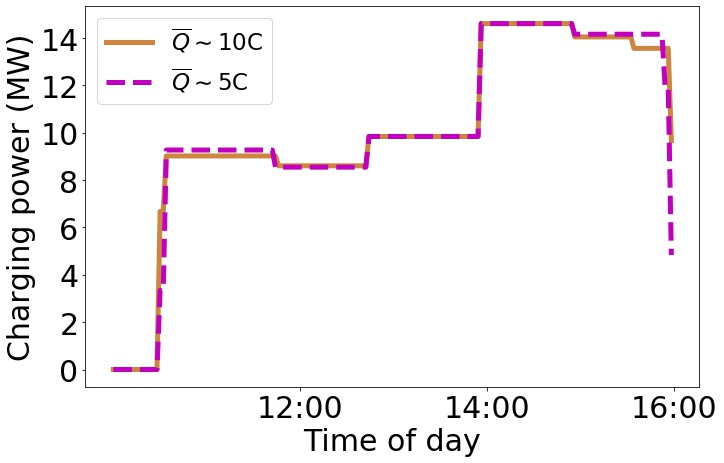}\label{fig:charging.2}}
  	\caption{Positive displacements and charging profiles with $\overline{P} = 20$ MW, $w = 1/\overline{P}^2$, $\overline{R} = 45$ and two different choices for battery charging capacities: $\overline{Q} \sim $ 10C and $\overline{Q} \sim $ 5C.}
\end{figure}

Next, we consider an upper bound of $\overline{P} = 20$ MW on the total power drawn by HEA at the JFK airport. This capacity is far less than the peak power of 35.9 MW obtained under a uniform charging schedule added to the rescheduling problem that ignores charging considerations. Encoding a realistic 10C battery charging rate in $\overline{Q}$'s, we run \eqref{eq:slot.alloc}  with two different choices of $w$. With $w=0$, we obtain a charging profile whose peak power is 20.0 MW, that equals the airport's charging capacity. That is, even without explicitly seeking to flatten the aggregate charging profile across HEA via the objective, the optimization problem finds a slot allocation and charging schedule that respects charging capacity limits at the airport, enforced via \eqref{eq:power.airport}. Even when the peak power respects said limits, it is useful to flatten the charging profile and reduce peak powers to avoid large peak demand charges. Upon choosing $w = {1}/{\overline{P}^2}$, we obtain a charging profile whose peak power is 14.6 MW, that is even less than that obtained with $w=0$. Thus, positive $w$ aids in peak shaving.\footnote{\rev{In general, it can happen that peak power drawn with $w=0$ and $w > 0$ become $\overline{P}$, where positive $w$ will flatten the resulting charging profile further than with zero $w$, albeit with coinciding peaks.} }
The displacements from these experiments are visualized in Figure \ref{fig:displacements.1} and the charging profiles are plotted in Figure \ref{fig:charging.1}. These figures demonstrate that charging considerations impact not only the charging schedules but they also affect the resulting flight schedules. In the same vein, changing the upper bound on the battery charging rate $\overline{Q}$ impacts both displacement decisions and charging profiles. Figures \ref{fig:displacements.2} and \ref{fig:charging.2} capture this sentiment through an experiment with $\overline{P} = 20$ MW, $w = 1/{\overline{P}^2}$, $\overline{R} = 45$, where $\overline{Q}$ encodes two different battery charging limits of 5C and 10C.

\begin{figure}[!hbtp]
  \centering   
    \subfloat[Histogram of positive displacements ]{\includegraphics[width=0.45\textwidth]{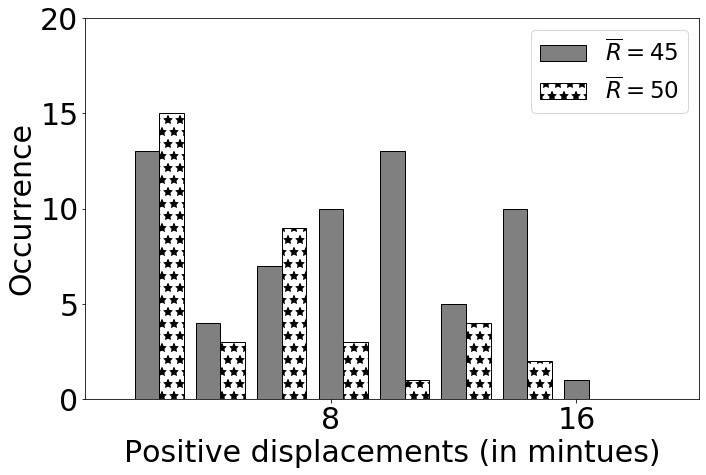}\label{fig:displacements.3}}
    \qquad
  	\subfloat[Charging profiles]{\includegraphics[width=0.465\textwidth]{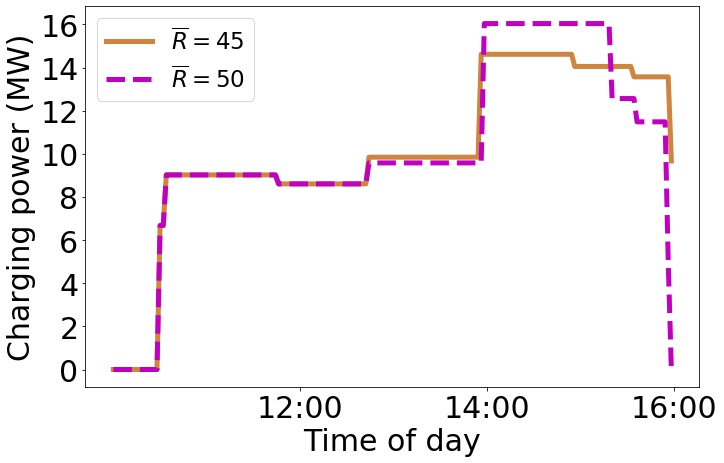}\label{fig:charging.3}}
  	\caption{Positive displacements and charging profiles for two different airport capacities ($\overline{R} = $ 45 and 50) with $\overline{P}= 20$ MW, $w = 1/\overline{P}^2$ and $\overline{Q} \sim 10$C.}
\end{figure}

As one might expect, expanding the airport capacity $\overline{R}$ to accommodate 50 flight movements instead of 45 movements per hour should reduce the extent of rescheduling. Indeed, aggregate displacements reduce non-linearly from 502 minutes with $\overline{R} = 45$ to 206 minutes with $\overline{R}=50$. Figure \ref{fig:displacements.3} confirms that the overall distribution of displacements skews leftward from this expanded runway capacity. This experiment utilizes $\overline{P} = 20$ MW, $w ={1}/ {\overline{P}^2}$ and $\overline{Q}$ encodes a charging rate upper bound of 10C. Change in runway capacity from $\overline{R} = 45$ to $\overline{R} = 50$ not only alters the flight schedules, but it also changes the resulting charging profiles (see Figure \ref{fig:charging.3}). The peak power changes from 14.6 to 16.0 MW. To explain this increase, note that an expanded airport capacity allows more arrivals and departures in each time interval. As the throughput during peak hours increases, the aggregate power demand from the HEA during these peak hours concomitantly increases. This experiment illustrates that airport's capacities to handle transportation throughput not only impact flight schedules but also significantly affect the charging profiles and their peaks. 

The above experiments demonstrate that constraints on airport capacities as well as those imposed on HEA charging affect both flight schedules and charging profiles with adoption of HEA in commercial aviation. One cannot simply tackle flight scheduling and HEA charging separately; these two questions are inextricably linked. We remark that while we only report the results from JFK, the framework is general and can be applied to any airport.

\rev{Finally, we consider the question of how proposed schedules are altered when the outputs of \eqref{eq:slot.alloc} cause longer displacements than what an airline deems acceptable (assume the maximal acceptable displacement is 20 minutes in this experiment for all airlines). To test the two mechanisms for altering proposed schedules, we first generate a schedule by running \eqref{eq:slot.alloc} with a declared capacity of $\overline{R}=60$, considering all aircraft to be conventional. In other words, we ignore all charging related variables in \eqref{eq:slot.alloc}. The solution features an aggregate displacement of 44 mins, with the maximal displacement of any one movement being 12 mins. The resulting schedule defines the proposed movements in our next experiment, but each airline offers to utilize HEA for as many aircraft routes that can be switched to HEA, per Section \ref{sec:energy}. As before, there are 32 HEA-operated connections at JFK within the considered time window, operated by 8 airlines.  We set $\overline{P} = 14.5$ MW and re-solve \eqref{eq:slot.alloc}. The altered schedule generates a total displacement of 144 mins with the maximal displacement being 116 mins.} 

\begin{figure}[!hbtp]
  \centering   
    \subfloat[Charging profiles ]{\includegraphics[width=0.43\textwidth]{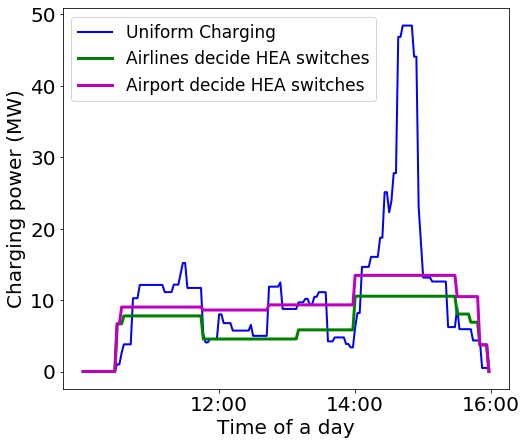}\label{fig:cl-charging}} 
    \qquad
  	\subfloat[Share of HEA operated flights]{\includegraphics[width=0.5\textwidth]{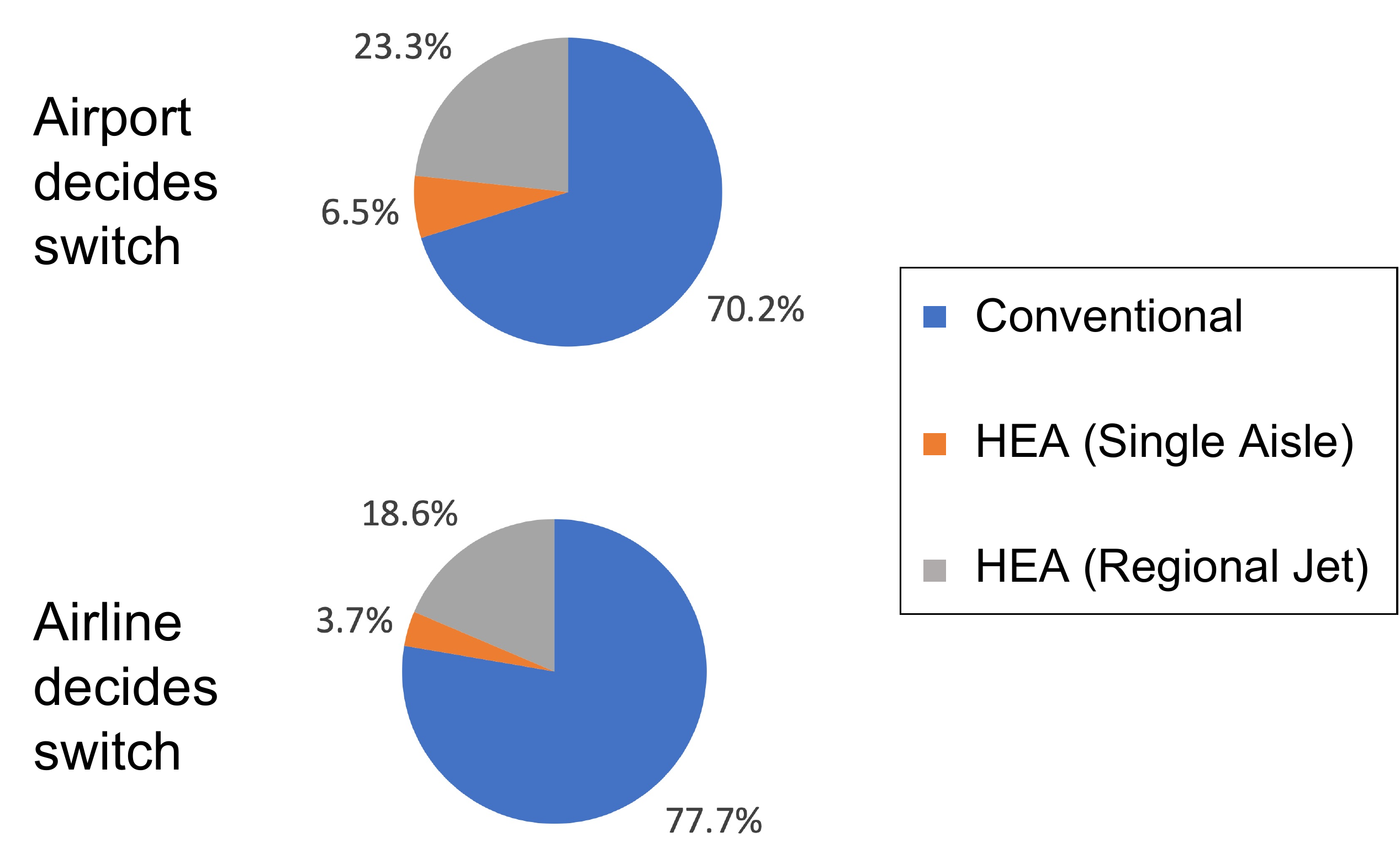}\label{fig:pie-before}}
  	\caption{Comparison of (a) charging profiles and (b) the share of HEA-operated aircraft routes under two different ways to switch such routes/paths to conventional aircraft.}
\end{figure}

\rev{Based on the modified schedule, we now consider the two schemes to switch HEA-operated connections to conventional aircraft. 
In the first scheme, the airport chooses to switch the single flight with maximum uniform charging rate among all HEA-operated connections (which also corresponds to the flight with maximal displacement). In the second scheme, each of the 8 airlines drops one HEA-operated connection, by switching those HEA-operated aircraft routes to conventional aircraft that require the maximum uniform charging rate among their movements. As a result, their switches convert 8 HEA-operated aircraft routes to conventional. While in general these two mechanisms will iterate, for this example, running \eqref{eq:slot.alloc} on the modified schedules produces no displacements, and hence, become acceptable to all airlines.
The resulting charging profiles for the two schemes are  portrayed in Figure \ref{fig:cl-charging}. The number of HEA-operated aircraft routes being lower with airlines making the switch in the second scheme, the energy requirements are naturally lower than that of the airport-controlled mechanism. The difference in the proportions of HEA-operated aircraft routes in Figure \ref{fig:pie-before} illustrates that practical considerations of congestion management can play a vital role in HEA use. Again, while we consider just two representative mechanisms for congestion management at JFK, our experimental setup is general, and can be applied to data from other airports with potentially other reschuduling schemes.}

\section{Conclusions and Future Directions}
\label{sec:conclusion}


HEA technology is maturing fast. They are projected to become viable for commercial aviation over the next few decades. While their overall energy needs at a national scale had been estimated before, we took a much more nuanced view of airport operations with HEA in this paper. Specifically, we provided a framework to gauge the energy needs of operating a specific aircraft's route with plausible hybrid electric options. This calculation allowed us to estimate the substantial increase in energy demands at major US airports with likely technology growth scenarios. 
\rev{Future technology growth patterns are inherently uncertain, and as a result, our projections of energy requirements bear the burden of that uncertainty. Given the variety of possible technology growth (BSED/MF combinations), our work provides an estimated range of possible energy capacities required at airports for HEA adoption.}
We showed through various examples that one must carefully design the charging profile of HEA at airports  to reduce peak power demands. Smart management of HEA charging profiles and slight alterations of flight schedules can help to significantly reduce peak power demands at airports. Such reductions can lighten the burdens of required grid infrastructure upgrades and allow airports to avoid peak electric demand charges. Building on this observation, we then proposed a flight rescheduling and charging algorithm that seeks to both minimize displacements of requested movements and flatten aggregate charging profiles. 
We illustrated our proposed formulation through a case study for JFK airport. The key insight from our analysis is that adoption of HEA within airline fleets will require coordination between flight scheduling and HEA charging. Scheduling and charging cannot be solved separately. \rev{While one can choose not to enforce flight rescheduling protocols at the cost of increased delays; by constrast, constraints imposed by the charging infrastructure cannot be frequently violated without seriously damaging the infrastructure. And coordination of scheduling and charging must be done \emph{across} airlines. Our results illustrate that the specifics of implementation, e.g., who decides HEA-operated flight paths, can substantially alter HEA usage, that in turn, will affect long-term HEA adoption decisions made by airline carriers.}

Our rescheduling algorithm with HEA charging is designed for a single airport. Such a framework can be extended in future work to jointly schedule flights and charge HEA across several airports. That framework will allow us to relax the requirement that each airport must  fulfill the charging needs of all its outgoing flights. Rather, one can charge HEA at only a few airports that upgrade their grid infrastructure. We do not anticipate conceptual difficulties in formulating such a problem. However, solving such an optimization problem at scale will invariably require careful algorithm design, \rev{especially when solved over a long time horizon such as 6 months}.
Note that our rescheduling and charging algorithm is meant as a planning tool that solves the problem prior to the date the flights are operated. Real-time contingencies such as weather-related variations in runway capacities, unexpected equipment malfunction and crew shortages at airports
inevitably require modifications of such plans. We plan to enhance our model to include tactical recourse decisions that adapt to said contingencies, \rev{possibly optimizing charging decisions with access to local stochastic solar power generation at an airport.} In this paper, we have not explicitly modeled the costs of HEA charging. In future work, we aim to study the design of contracts among electric utilities, airports and airlines to pay for powering HEA. Such a study will allow us to estimate how HEA adoption will impact flight ticket prices as such costs trickle down to passengers.




\appendix
\section{Characteristics of HEA Concepts}
\label{sec:supp}
\setlength{\tabcolsep}{5pt}
\renewcommand{\tablename}{Table}
\setcounter{table}{0} 
\begin{table}[H]
\caption{Battery energy usage $b_0$ in Wh per passenger-mile}
\label{tab:Batt}
\centering
\begin{tabular}{cccccccc}
\toprule
\multirow{2}{*}{Type} & \multirow{2}{*}{MF (\%)} &\multicolumn{5}{c}{BSED (Wh/kg)} \\\cmidrule(lr{1em}){3-7}& &  500 &   700 &     1000 &     1250 &     1500 \\
\midrule
\multirow{3}{*}{Boeing 737-700} &12.5 &  26.5 &   26.3 &  25.7 &  25.4 &   25.2 \\
				  &25   &  57.3 & 55.5 &  54.0 &  53.2 &   52.6 \\
				  &50   &     - &  119.8 &  115.1 &  112.8 &  111.2 \\
\midrule
\multirow{3}{*}{ERJ-175}&12.5 &  31.1 &  30.7 &  30.4 &   30.2 &   30.0 \\
			 	   & 25   &  64.8 &  63.6 &  62.6&   62.1 &   61.6 \\
			  	   & 50   &     - & 133.3 &  130.6 &  129.1 &  127.9 \\
\bottomrule
\end{tabular}

\vspace{0.5cm}

\caption{Maximum range of HEA in miles} 
\label{tab:Range} 
\centering
\begin{tabular}{cccccccc}
\toprule
\multirow{2}{*}{Type} & \multirow{2}{*}{MF (\%)} &\multicolumn{5}{c}{BSED (Wh/kg)} \\\cmidrule(lr{1em}){3-7}& &500 &      700 &     1000 &     1250 &     1500 \\
\midrule
\multirow{3}{*}{Boeing 737-700} & 12.5 &    1110.9 &  1466.2 &   1876.8 &   2141.3 &   2357.5 \\
				        &  25   &       558.9 &   836.1 &   1193.7 &  1452.4 &  1680.1\\
				        & 50   &     - &   376.1 &    637.1 &   842.9 &  1038.4\\
\midrule
\multirow{3}{*}{ERJ-175}& 12.5 &   683.1 &   914.2 &   1170.7 &  1339.7 &   1476.6 \\
			       &  25   &    305.9 &   484.1 &   714.1 &   877.45&  1020.1 \\
			       &  50   &   - &    170.2 &    331.2 &   458.8 &    579.6 \\ 
\bottomrule
\end{tabular}
\end{table}

\section{Number of Flights in 2018 BTS Data Within HEA Range Capabilities with Various BSED-MF Configurations}
\label{sec:supp.flight}

\begin{table}[H]
\caption{Number of flight paths from the various airports in 2018, that were operated by a regional jet (RJ) or single aisle (SA) aircraft in the BTS dataset.}
\label{tab:hybrid-supp}
\centering
\begin{tabular}{cc}
\toprule
Airport & \# flights served by RJ/SA \\
\midrule
ATL & 289,951 \\
ORD & 304,651 \\
DFW & 239,156 \\
SFO & 132,321 \\
IAD & 61,455 \\
SAN & 56,729 \\
\bottomrule
\end{tabular}
\end{table}

\label{sec:supp-figure1}
\begin{figure}[hbtp!]
  \centering
  	\subfloat[ATL]{\includegraphics[width=0.3\textwidth]{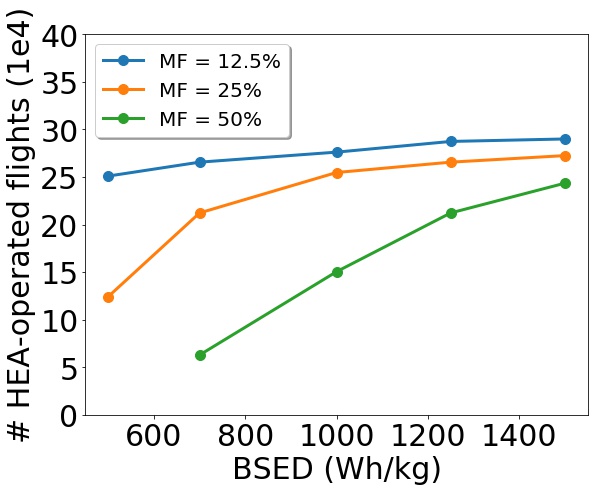}}
	\subfloat[ORD]{\includegraphics[width=0.3\textwidth]{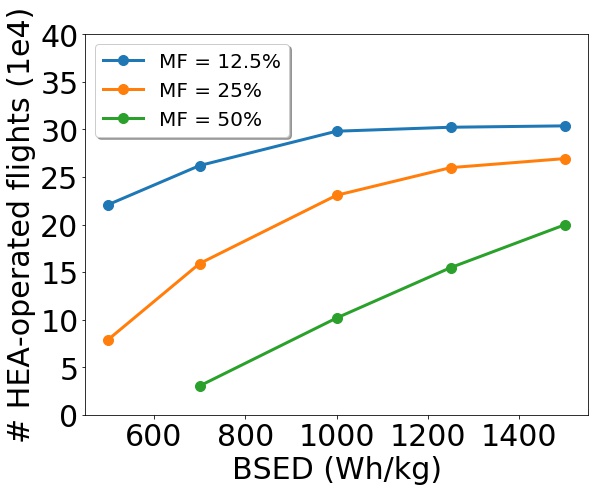}}
	\subfloat[DFW]{\includegraphics[width=0.3\textwidth]{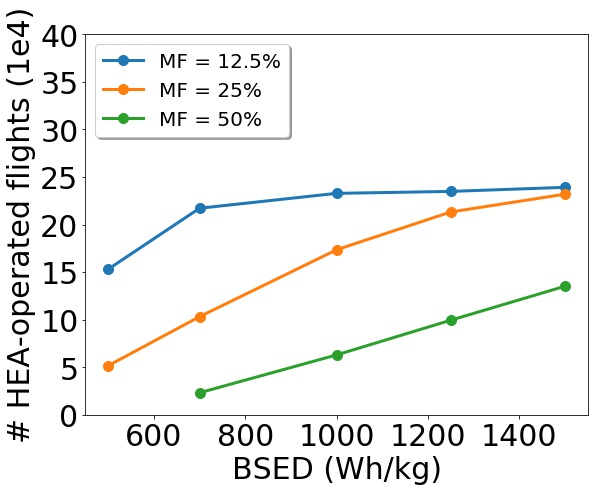}}
	\vspace{0.1in}
	\subfloat[SFO]{\includegraphics[width=0.3\textwidth]{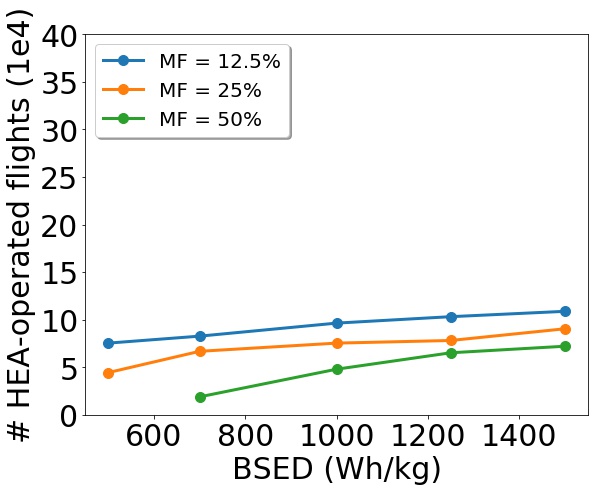}\label{fig:hybrid.SFO}}
	\subfloat[IAD]{\includegraphics[width=0.3\textwidth]{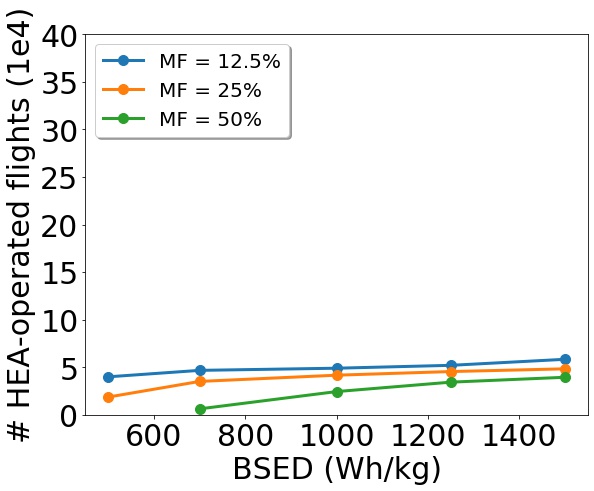}}
	\subfloat[SAN]{\includegraphics[width=0.3\textwidth]{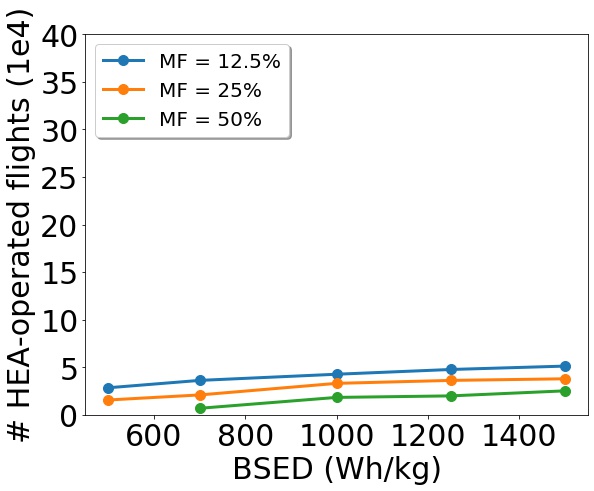}}
	\caption{Number of flights that are switched to HEA under various BSED/MF combinations.}
  	\label{fig:hybrid-supp}
\end{figure}

\bibliographystyle{elsarticle-num-name} 
\bibliography{elsarticle.bib}

\end{document}